\documentstyle[sprocl]{article}

\def\Journal#1#2#3#4{{#1} {  #2}, #3 (#4)}

\def\NCA{{\em Nuovo Cimento} A}

\def\NPB{{\em Nucl. Phys.} B}
\def\PLB{{\em Phys. Lett.}  B}
\def\PRL{\em Phys. Rev. Lett.}
\def\PRD{{\em Phys. Rev.} D}
\def\ZPC{{\em Z. Phys.} C}
\def\NPA{{\em Nucl. Phys.} A}

\def\ZPA{{\em Z. Phys.} A}

\def\be{\begin{equation}}
\def\ee{\end{equation}}
\def\bea{\begin{eqnarray}}
\def\eea{\end{eqnarray}}

\begin{document}

\pagestyle{empty}
\vskip.9cm       
\title{  Study of Multiquarks Systems in a Chiral Quark Model} 
\vskip .7 cm                                                                    
\author{\underline{M. GENOVESE} \footnote{ \small  Supported by the EU Program ERBFMBICT 950427}
 and  J.-M. RICHARD }
\address{Institut des Sciences Nucl\'eaires \\
Universit\'e Joseph Fourier--IN2P3-CNRS, \\
53, avenue des Martyrs, F-38026 Grenoble Cedex,  
France}                                                             
\vskip .2cm                                               
\author{S. PEPIN and   Fl. STANCU}
\address{Universit\'e de Li\`ege \\ Institut de Physique, 
B.5, Sart Tilman, \\ B--4000 Li\`ege 1, Belgium}             
                                                                    
\vskip .5cm                                                         
\maketitle\abstracts{We  discuss the stability of  multiquark systems within the recent model
of Glozman {\sl et al.\/} where the chromomagnetic hyperfine interaction
is replaced by  pseudoscalar-meson exchange contributions. 
In this model the ($u,d$) diquark $S=0$ $I=0$ is strongly bound 
(more than for the chromomagnetic interaction) and this leads to
a bound heavy tetraquark $QQ\bar q\bar q$ where the light quarks 
are in the $S=0$ $I=0$ configuration. We study the stability of other multiquark systems 
as well.}

\section{Introduction}

\noindent The interest for the possible existence of 
multiquark hadrons has been raised twenty years ago by Jaffe, 
who suggested that states of two quarks - two antiquarks \cite{1} and
of six quarks \cite{J2} could be bound.

In the following years this problem has been studied within a 
large variety of models. Some earlier studies in MIT bag 
indicated the presence of a dense spectrum of tetraquark states in
the light sector \cite{1} (and more generally of multiquarks 
\cite{aerts}). Later on, tetraquark systems have been
examined in potential models \cite{2,3,4,5} and flux tube
models \cite{6}.  Weinstein and Isgur
showed \cite{2} that there are only a few weakly bound states of
resonant meson--meson structure in the light
$(u,\,d,\,s)$ sector. Usually the ground state 
of these systems lies closely above
the lowest $(q\bar q)+(q\bar q)$ threshold. 
The other extreme result, as compared to MIT bag model, was obtained by
Carlson and Pandharipande \cite{6} in their flux-tube model with
quarks of equal masses, where no bound state was found.%

Altogether the theoretical predictions about the existence of these bound
states are still unclear.

As far as the experimental situation is concerned, 
there are some candidates for non-$q\bar q$
states, but data are not yet conclusive. Furthermore, even if a resonance
would be clearly identified as an exotic (not $q \bar q$ or $qqq$) state,
nevertheless a careful phenomenological study would be necessary in order
to understand its internal structure among many different possibilities
(multiquarks, hybrids, glueballs,...).

The description of these candidates and of the phenomenological properties
which permit a distinction between different exotics is beyond the purposes of
this proceeding and we refer to  Ref.s~\cite{7,NC} and the last issue of Review of
Particle Properties \cite{8} for further details.

Anyway, a well-established theoretical result is that 
systems with a larger mass difference among their components
are more easily bound \cite{R3,3}.
For example, for a
system of two heavy quarks and two light antiquarks
$QQ\bar q\bar q$ ($Q=c$ or $b$, $q=u,\,d$ or $s$) stability can be
achieved without spin--spin interaction, provided the mass ratio
$m(Q)/m(q)$ is larger \cite{3} than about 15, which means that $Q$
must be a $b$-quark.

Recently, the interest for multiquark systems containing charm 
quarks has grown, considering that new experiments are
being planned at Fermilab and CERN, to search for new  hadrons  and
in particular for doubly charmed tetra\-quarks \cite{9,10,11}.

In this context, we have carried out \cite{nos1} a study of
the $QQ \bar q \bar q$
system in the framework of the chiral quark model of Glozman {\sl et
al.\/} \cite{12,13}. This model is somehow quite ``extreme'' because it includes
meson-exchange forces between quarks and entirely neglects the
chromomagnetic interaction. However it permits a very good description
of the baryon spectrum, it is thus worth testing it in
further predictions. 

Considering that in this model light-quark mesons are 
``quasiparticles''
with a spectrum which must be assumed and cannot be evaluated directly
(this is of course rather an unpleasant feature of the model), multiquark
systems represent an obvious testing ground for it.   
Other possible tests could come from a careful analysis of the
 spectrum of charmed baryons \cite{nos}.

\section{The Glozman model}

Before presenting our results about multiquarks, let us briefly
consider the Glozman {\it et al.} model and compare it with other potential models
used in hadron spectroscopy.

In a general Hamiltonian, which would approximate the low energy limit 
of QCD, one can introduce both a chromomagnetic interaction and
a me\-son-ex\-chan\-ge contribution, obtaining an explicit form as
$$
 H=\sum_i{\vec{\rm p}_i^{\,2}\over 2m_i}-{3\over16}                            
    \sum_{i<j} \tilde{\lambda}_i^{\rm c}\!\cdot\!
\tilde{\lambda}_j^{\rm c}\,                                          
               V_{\rm conf}(r_{ij})
$$
$$
\,\,\,
{}-\sum_{i<j}\tilde{\lambda}_i^{\rm                                            
c}\!\cdot\!\tilde{\lambda}_j^{\rm c}\,                                          
\vec{\sigma}_i\!\cdot\!\vec{\sigma}_j\,V_{\rm g}(r_{ij})-                       
\sum_{i<j}\tilde{\lambda}_i^{\rm F}\!\cdot\! \tilde{\lambda}_j^{\rm             
F}\,                                                                            
\vec{\sigma}_i\!\cdot\!\vec{\sigma}_j \,V_{\rm F}(r_{ij}),                      
\eqno(1) 
$$                                                                            
where $m_i$ is the constituent mass of the quark located at
$\vec{\rm r}_i$; $r_{ij}=\vert \vec{\rm r}_j-\vec{\rm r}_i\vert$
denotes the interquark distance; $\vec{\sigma_i}$,
$\tilde{\lambda}_i^{\rm c}$,
$\tilde{\lambda}_i^{\rm F}$ are the spin, colour and
flavour  operators, respectively. Spin-orbit and
tensor components may supplement the above spin-spin forces for
studying orbital excitations (they give no contribution for $L=0$ systems). 
The potential in (1) has three parts
containing the confining, the chromomagnetic and the meson-exchange
contribution.

Usually, the confining term $V_{\rm conf}$ is assumed to include a Coulomb plus
a linear term,                               
$$                                                                              
V_{\rm conf}=-{a\over r}+b r+c.\eqno(2)                                           
$$
In the following, we shall either use  the very weak linear potential
of Glozman {\sl et al.\/} \cite{13} corresponding to
$$
(C_1)\qquad a=c=0, \qquad\hbox{and}\qquad b=0.01839\;{\mathrm GeV}^2,
\eqno(3)
$$
or the more conventional  choice
$$
(C_2)\qquad a=0.5203, \qquad b=0.1857\;{\mathrm GeV}^2,
\qquad c=-0.9135\;{\mathrm GeV},\eqno(4)
$$
which has already been applied to the study of tetraquarks by
Silvestre-Brac and Semay \cite{4}.%

The term $\sum_{i<j}\tilde{\lambda}_i^{\rm                                            
c}\!\cdot\!\tilde{\lambda}_j^{\rm c}\,                                          
\vec{\sigma}_i\!\cdot\!\vec{\sigma}_j\,V_{\rm g}(r_{ij})$
is the chromomagnetic
analogue of the Breit--Fermi term of QED. In the interaction
between a quark and an antiquark (as e.g. in a meson) one finds 
$\tilde{\lambda}_1^{\rm c}\!\cdot\!\tilde{\lambda}_2^{\rm c}=-16/3$.
Then a positive $V_{\rm g}$ shifts
each vector meson above its  pseudoscalar partner, for instance
$D^*>D$ in the charm sector. For baryons, where
$\tilde{\lambda}_1^{\rm c}
\!\cdot\! \tilde{\lambda}_2^{\rm c}=-8/3$ for each quark pair,  such
a positive $V_{\rm  g}$ pushes the spin 3/2 ground states up, and the
spin 1/2 down, for instance $\Delta >N$. For the radial 
shape, as an example, we mention 
$$
V_{\rm g}={a\over\mathstrut m_im_jd^2}
{\exp-r/d\over\mathstrut r},\eqno(5) 
$$
which was used in Ref.\ \cite{4}, 
with the same value of $a$ as in Eq.~(4) and $d=0.454\;$GeV$^{-1}$.
                  
Finally, the last term of $H$ corresponds to meson exchange, and an explicit
sum over $F$ is understood. If the system contains light quarks only (as in 
Ref.s \cite{12,13}),
the sum over $F$ runs from 0 to 8, {\sl i.e.} over the members of 
the $J^{PC}=0^{-+}$ nonet,
which represent the Goldstone bosons of the 
spontaneusly broken $SU(3)_{A}$ symmetry
 ($1-3\rightarrow\pi$, $4-7\rightarrow
K$, $8\rightarrow\eta$  and  $0\rightarrow\eta'$). If a heavy flavour
is incorporated,  a phenomenological extension from 
SU(3)$_{\rm F}$ to
SU(4)$_{\rm F}$ would further extend the sum to  ${\rm F} = 9-12$
corresponding to a $D$-exchange, ${\rm F} = 13-14$ to a
$D_s$-exchange and ${\rm F} = 15$ to an $\eta_c$-exchange
(of course in this case the interpretation as Goldstone bosons is 
not really possible). Similar terms
should then be introduced if one includes the beauty sector as well. 
The radial form of $V_{\rm F}\neq0$ is
derived from the usual  pion-exchange potential which contains a
long-range part and a short-range one
$$                                                                              
\sum_{i<j}\vec{\tau}_i\!\cdot\!\vec{\tau}_j\,\vec{\sigma}_i\!\cdot\!
\vec{\sigma}_j                                                                    
{g^2\over 4\pi}{1\over 4m^2}\left[\mu^2{\exp(-\mu r_{ij})\over                  
r_{ij}}-4\pi\delta^{(3)}(r_{ij})\right],\eqno(6)                                
$$
where $\mu$ is the pion mass. 
                                                                                
When constructing $N\!N$ forces from meson exchanges, one 
usually disregards 
the short-range term in Eq.~(6), for it is hidden by the hard core, 
and anyhow the potential in that region is parameterized empirically. 
For example, when  T{\"o}rnqvist \cite{14}, Manohar and
Wise \cite{15} or Ericson and Karl \cite{16}
considered pion exchange in multiquark states, they used
the Yukawa term $\exp(-\mu r)/r$ acting between two well-separated quark
clusters. Weber {\sl et al.\/} \cite{17},  in
their model with hyperfine plus pion-exchange interaction, studied both the
cases with or without delta-term, showing
that good results could be obtained
without it in baryon spectroscopy.
Thus the relevance of an {\sl ad-hoc\/} regularized delta-term 
\cite{12,13} in the study of baryon spectroscopy is somehow surprising.
Nevertheless, this ansatz permits to give a good description of 
the baryon spectrum and, in particular, allows one to solve the problem
of the ordering of the lowest  parity-odd and parity-even states
of $N$, $\Lambda$ and $\Sigma$ resonances, which did not find a
solution in conventional chromomagnetic models \cite{Chromo}.

Incidentally, this result is not completely unexpected, in fact 
Buchman {\it et al.}
had shown, already some years ago, 
that this term is essential for obtaining a good description of
magnetic moments \cite{buch}.

For the sake of completeness we report here the explicit form
of the regularized delta-term \cite{13}
$$
V_\mu=\Theta(r-r_0)\mu^2{\exp(-\mu r)\over
r}-                                 
{4\epsilon^3\over\sqrt\pi}\exp(-\epsilon^2(r-r_0)^2),                                    
\eqno(7)
$$
where 
$r_0=2.18\;$GeV$^{-1}$, $\epsilon=0.573\;$GeV, and
$\mu= 0.139$ GeV for $\pi$, $0.547$ GeV for $\eta$ and $0.958$ GeV for 
$\eta '$. Furthermore, the Yukawa-type part is cut off for $r\le r_0$.
This smearing should account for
the fact that both the pseudoscalar mesons and the constituent quarks
have a finite size and that boson fields cannot be described by 
a linear equation near their source.

The explicit form of the Hamiltonian
which extends the results of Ref. \cite{12} from SU(3) to SU(4)
is given in Ref. \cite{nos1}.
It includes also the exchange of $D$, $D_s$ and $\eta_{c}$ mesons.
However, considering that little
$u\bar u$ or $d\bar d$ mixing is expected in $\eta_c$, this
contribution can be 
neglected when considering systems with none or one charm quark.
Moreover when the meson mass $\mu$ reaches
values of a few GeV as for $D$ or $\eta_c$ the two terms in Eq.~(6)
basically cancel each other and one recovers practically the 
SU(3) form \cite{12}.  This is in agreement
with Ref.~\cite{18} where it has been explicitly shown that the dominant
contribution to the $\Sigma_c$ and $\Sigma_c^*$ masses is due to
meson exchange between light quarks and the contribution of the
matrix elements with the $D$ ($D_s$) and $D^*$ ($D^*_s$) 
quantum numbers (which are evaluated phenomenologically, 
fitting the mass difference $\Sigma_c - \Lambda_c$) 
play a minor r\^ole. In the following
numerical calculations we will neglect the exchange of heavy mesons.%

Finally one has to fix quark masses. The light quarks ones are fixed to 
$0.34$ GeV according to Ref.s \cite{13,4}.
The heavy quark masses $m_Q=m_c$ and $m_b$ are adjusted to reproduce the
experimental average mass $\overline{M}=(M+3M^*)/4$ between the 
$0^{-}$ and $1^{-}$ $M=D$ or
$B$ mesons by a variational calculation, where a trial wave function of
type $\phi\propto\exp(-\alpha r^2/2)$ is used (with $\alpha$ as a variational
parameter). It has been checked that the error never exceeds a few
MeV with respect to the exact value. The variational approximation
is retained for consistency with the treatment of 3-, 4- and 6-body
systems discussed below. This leads to $m_c=1.35$ GeV and $m_b=4.66$ GeV 
for the potential $C_1$ and $m_c=1.87$ GeV and $m_b=5.259$ GeV
for $C_2$.

Before proceeding further, let us briefly discuss the calculation of 
baryons masses in the model of Glozman {\it et
al.\/} The explicit form of the Hamiltonian integrated in the
spin--flavour space is :
$$                                                                              
H=H_0+{g^2\over48\pi m^2}                                                       
\left\{                                                                         
\normalbaselineskip=20pt                                                        
\matrix{                                                                        
&\!\!\!\!\!15V_\pi-V_\eta-2\left({g_0/                                          
g}\right)^2V_{\eta'}\quad\hbox{for}\quad
N\cr                                                                                                                   
&\!\!\!\!\!\phantom{1}3V_\pi+V_\eta+2\left({g_0/g}\right)^2V_{\eta'}
\quad\hbox{for}\quad\Delta\cr                                                     
}\right.                                                                                                                                                
\eqno(8)$$                                                                              
with                                                                            
$$                                                                     
H_0= 3m+\sum_i{\vec{\rm
p}_i^{\,2}\over                                 
2m}+{b\over2}\sum_{i<j}r_{ij},                                                                                            
\eqno(9)                                                                        
$$ 
where $g^2/4\pi = 0.67$ (which leads to the usual strength
$g_{\pi N\!N}/4\pi\simeq14$ for the Yukawa
tail of the nucleon--nucleon potential) and $(g_0/g)^2 = 1.8$.
We have performed variational estimates with a wave
function $\phi \propto \exp(-\alpha (\rho^2 + \lambda^2)/2)$,
where $\vec{\rho}= \vec{\rm r}_2  - \vec{\rm r}_3$,
$\vec{\lambda}=(2\vec{\rm r}_1-\vec{\rm r}_2-\vec{\rm r}_3)/\sqrt{3}$:
our results agree with  the more elaborated Faddeev
calculations of Ref.~\cite{13}.

When the meson--exchange terms are switched
off, the $N$ and $\Delta$ ground states are degenerate at
1.63 GeV. Introducing the coupling  the nucleon mass drops
stronger leading to a reasonable splitting ($\approx 0.3\;$GeV).
It is interesting to notice that in this model one has an attraction both 
for the $S=0$, $I=0$ and also, albeit smaller, for the $S=1$, $I=1$ 
diquarks. Then both the nucleon and $\Delta$ masses decrease when the 
interaction is turned on. The diquarks $S=0$, $I=1$ and $S=1$, $I=0$
on the contrary are repulsive configurations. The situation is thus 
different from the chromomagnetic case where one has a smaller 
attraction for the $S=0$, $I=0$ diquark and repulsion for the $S=1$, 
$I=1$ one.

We have also calculated the ground state of $cqq$ baryons using
a trial wave function $\phi \propto 
\exp(-(\alpha \rho^2+\beta\lambda^2)/2)$ and found 
$\Lambda_{c} = 2.32\;$GeV and $\Sigma_c = \Sigma^{*}_{c} =
2.48\;$GeV, close to the experimental values and consistent with the
findings of Ref.~\cite{18}, although the Hamiltonian, its treatment,
and the input parameters are somewhat different there.
   
\section{Evaluation of multiquarks masses}

Due to arguments at the beginning of this contribution, here we discuss
tetra\-quarks containing heavy flavours, i.e. $QQ \bar q \bar q$, 
studying the most favourable
configuration:  $\bar 3 3,\,S=1,\,I=0$. This means that $QQ$ is
in a  $\bar 3$ colour state and $\bar q\bar q$ in a 3 colour state.
The mixing with $6\bar6$ is neglected because one expects 
this to play a negligible r\^ole in deeply-bound heavy
systems \cite{3}. Then the Pauli principle requires $S_{12}=1$ for
$QQ$, and $S_{34}=0,\, I_{34}=0$ for $\bar q\bar q$ (which we have seen
to be the diquark with the largest binding), if the relative
angular momenta are zero for both subsystems. This gives a state of
total spin $S=1$ and isospin $I=0$. 

The tetraquark Hamiltonian integrated in the colour--spin--flavour
space, and incorporating the approximations discussed in the former 
section, reduces to
$$H=2(m+m_Q)+{\vec{\rm p}_x^2\over m_Q}+{\vec{\rm p}_y^2\over m}
+{m+m_Q\over2mm_Q}\vec{\rm p}_z^2+\sum_{i<j}V_{ij},\eqno(10)
$$
where
$$
V_{12}={1\over2}\left(-{a\over r_{12}}+b\, r_{12}+c\right),
$$
$$
V_{ij}={1\over4}\left(-{a\over r_{ij}}+b\, r_{ij}+c\right),
\qquad i=1\;\hbox{or}\;2,\;j=3\;\hbox{or}\;4,
\eqno(11)
$$
$$
V_{34}={1\over2}\left(-{a\over r_{34}}+b\, r_{34}+c\right)
+9V_\pi-V_\eta-2V_{\eta'}.
$$

The momenta $\vec{\rm p}_x$, etc., are conjugate to the relative
distances $\vec{\rm x}=\vec{\rm r}_1-\vec{\rm r}_2$, 
$\vec{\rm y}=\vec{\rm r}_3-\vec{\rm r}_4$, and 
$\vec{\rm z}=(\vec{\rm r}_1+\vec{\rm r}_2-\vec{\rm r}_3-\vec{\rm
r}_4)/\sqrt2$. The wave function is parameterized as
$$
\psi\propto\exp[-(\alpha x^2 +\beta y^2+\gamma z^2)/2],
\eqno(12)
$$
and the minimization with respect to $\alpha,\,\beta$ and $\gamma$ 
shows that both the $cc \bar q \bar q $ and $bb \bar q\bar q$ systems
are bound whatever is the potential, $(C_1)$ or $(C_2)$, provided
meson exchange is incorporated. This is in contradistinction to
previous studies based on conventional models where the flavour-independent 
confining potential is supplemented by one gluon exchange. 
For example the authors of 
Ref.\cite{4} found that the $cc \bar q \bar q$ state is 
about $20$MeV above threshold (while the $bb \bar q \bar q $ is bound of 
$135$ MeV). A similar situation is also obtained in Ref. \cite{Don}, where
studying the four quarks  states in a diquark model only
the $b b \bar q \bar q $ is found to be bound (by $50$MeV, while $cc 
\bar q \bar q$ is $30$ MeV above the threshold).

Quantitatively  in our model we find that using the confining potential 
$C_1$ plus meson exchange the double charmed tetraquark is bound by
$185$ MeV and the double bottom one by $226$ MeV. Using the potential
$C_2$ the results are nearly two times larger ($332$  and $497$ MeV 
respectively) and also much more different from each other.
The reason is that ($C_2$) contains a Coulomb part which binds more, heavier is
the system, leading thus to a larger separation among levels as well.
This is related to the fact that
the potential ($C_2$) has been fitted to reproduce the $J/\Psi$ and the
$\Upsilon$ meson masses (anyway it also gives overall good results 
both for other mesons and baryons) while, by construction \cite{12,13},
the potential ($C_1$) was designed
and fitted to light baryons only.
Altogether, our results shows that the prediction of binding
for the $c c \bar q \bar q$ is substantially independent on the choice 
of the confining potential.

Considering this result one could rise the question if in the model
of Glozman  {\sl et al.\/} a
proliferation of multiquark systems appears. We have therefore tried
to investigate $QQqqqq$ and $q^6$ systems as well. 

As a general procedure, for a given multiquark system,
one searches for the  spin-isospin wave functions corresponding to 
a colour singlet, then selects the most favourable configuration.
The contribution of global spin-flavour-averaged interaction is 
reduced to the calculation of the matrix elements of the two body
operator $(\vec{\sigma}_i\!\cdot\!\vec{\sigma}_j) 
(\vec{\tau}_i\!\cdot\!\vec{\tau}_j)$; this is accomplished by
using Clebsh--Gordan coefficients of the permutation group
according to Ref.s \cite{19,PS}.
   
Let us begin with $QQqqqq$. In this case the most favourable 
configuration is the one where the light quark subsystem
has $S=1$, $I=0$, which leads to the  spin-flavour-averaged interaction
$$
\langle V \rangle=10 V_{\pi} - 2/3 V_{\eta} - 4/3(g_0/g)^2
V_{\eta '}.
\eqno(13)
$$

In the numerical calculation we have used the variational
Gaussian wave function 
$$
\Psi \propto \exp [-(\alpha x^2 + \beta y^2 + \gamma (u^2 + v^2 + w^2)]
\eqno(14)
$$
where appear the Jacobi variables
$\vec{\rm x}=\vec{\rm r}_1-\vec{\rm r}_2$,
$\vec{\rm y}=\vec{\rm R}_q-\vec{\rm R}_Q$,
in terms of $\vec{\rm R_q}=(\vec{\rm r}_1+\vec{\rm r}_2)/2$ and
$\vec{\rm R_Q}=(\vec{\rm r}_3+\vec{\rm r}_4+\vec{\rm r}_5+\vec{\rm r}_6)/4$,
$\vec{\rm u}=(\vec{\rm r}_3+\vec{\rm r}_4-\vec{\rm r}_5-\vec{\rm r}_6)
/\sqrt{2}$,
$\vec{\rm v}=(\vec{\rm r}_3-\vec{\rm r}_4+\vec{\rm r}_5-\vec{\rm r}_6)
/\sqrt{2}$,
$\vec{\rm w}=(\vec{\rm r}_3-\vec{\rm r}_4-\vec{\rm r}_5+\vec{\rm r}_6)
/\sqrt{2}$, where indices $1,2$ refer to the heavy quarks and $3,4,5,6$ to the
light ones.
 
The result of the numerical calculation is that this potential is largely
insufficient to bind both the $cc q qqq $ ({\sl e.g.} of $500$ MeV with the 
potential $C_1$) and  the 
$b b qqq q$ systems. Incidentally, at this workshop results about
heavy hexaquarks were presented also by Lichtenberg {\sl et al.} 
\cite{Don2}, 
in their diquark model. Albeit they do not treat directly the case
$cc qqqq$, they find that the system $csqqqq$ is unbound, while the one
containing a beauty quark instead of the charm is bound.

For the sake of completeness we have also made a study of  the $q^6$ system.
The most favourable
configuration is $S=1$, $I=0$, leading to 
$$
 \langle V \rangle=
11 V_{\pi} - 5/3 V_{\eta} - 10/3(g_0/g)^2
V_{\eta '}
\eqno(15)
$$
Our result shows that, also in this case, 
the system is largely insufficiently bound  (by almost $1.5$GeV with 
the potential $C_1$) for being under 
the two baryons threshold.

\section{Conclusions}
                                          
In conclusion, considering the success of the Glozman {\sl et al.} model 
in describing the baryon spectrum, we have searched for
new possible tests of 
this model. Unluckily the model does not permit an analysis of mesons 
(light-quark mesons at least), which are interpreted as 
``quasiparticles'' related to the breaking of
flavour $SU(3)$. Further tests of it would then involve multiquarks 
systems.
 We have thus studied four and six-quark systems involving two 
heavy quarks, showing that  the Glozman {\sl et al.} model leads to predictions 
 for the $c c \bar q \bar q$ which differ from the ones of more
 conventional models.
In fact this system is found to be strongly bound. The search of such a
resonance in 
forthcoming experiments will therefore be a good test  for understanding 
the dynamics which produces the hadron spectrum.

We have also found that the $b b \bar q \bar q$ state is bound, but this result
is obtained also in other models and the possibility to observe this resonance
is relegated to a more remote future.

Finally we have found that the six quark system with two heavy quarks is 
unbound, also this result does not differ from the one of other more 
conventional models. However, together with the outcome that also the
$q^6 $ system is unbound, it is a useful indication that in the 
framework of the model under 
investigation there is no proliferation of multiquark bound states.

\section*{Acknowledgements} 
We thank D.B. Lichtenberg for enlighting discussions and the organizers
for the stimulating atmosphere of this workshop. 

\vskip 1cm


\begin{thebibliography}{99}

\bibitem{1}R.L. Jaffe,
\Journal{\PRD} {15} {267} {1977}; \Journal{\PRD} {17} {1444} {1978}.
%
\bibitem{J2}R.L. Jaffe, \Journal{\PRL} {38} {195} {1977}.

\bibitem{aerts} A. T. M. Aerts {\sl et al.}, \Journal{\PRD} 
{17} {260} {1977}.

\bibitem{2}J. Weinstein and N. Isgur, \Journal{\PRD} {27} {588} 
{1983}, \Journal{\PRD} 
{41} {2236} {1990}.
%
\bibitem{3}S.Zouzou, B. Silvestre-Brac, C. Gignoux and J.-M.                     
Richard, \Journal{\ZPC} {30} {457} {1986}.
%
\bibitem{4}B. Silvestre Brac and C. Semay, \Journal{\ZPC} {59} {457}
{1993)}; \Journal{\ZPC}{61} {271} {1994}.
%
\bibitem{5} D.M. Brink and Fl. Stancu, \Journal{\PRD} {49} {4665}
{1994}.              

\bibitem{6}J. Carlson and V.R. Pandharipande,\Journal{ \PRD}
 {43} {1652} {1991}.
%
\bibitem{R3} J.P. Ader, J.-M. Richard and P. Taxil, 
\Journal{\PRD} {25} {2370} {1982}.

%
\bibitem{7}{G. Karl},
\Journal{{\em Int. J. of Mod. Phys.}} {E1} {491} {1992};
\Journal{\NPA} {558} {113c }{1993};
{N.A. T\"ornqvist},
{Proc. of ``Int. Europh. Conf. on High Energy Phys.'',
Brussels (Belgium), edit. J. Lemonne et al., 84 (1995);} 
{G. Landsberg}, \Journal{{\em Phys. of Atom. Nucl.}} {57} {42 } { 1994}.
%
\bibitem{NC} M. Genovese, \Journal{\NCA} {107} {1249} {1994} and 
references therein. 

\bibitem{8}Particle Data Group, \Journal{\PRD} {54} {1} {1996}.
%
\bibitem{9}M.A. Moinester,
\Journal{\ZPA} {355} {349} {1996}.
%
\bibitem{10}D.M. Kaplan,
Proc. Int. Workshop ``Production and Decay of Hyperons, 
Charm and Beauty Hadrons'',
Strasbourg (France) September 5--8, 1995.
%
\bibitem{11}COMPASS Collaboration (G. Baum et al.),
CERN--SPSLC--96-14, March 1996.
%
\bibitem{nos1} S. Pepin, Fl. Stancu, M. Genovese and J.-M.Richard,
hep-ph 9609348, ISN-96.99, to be published in \PLB .


\bibitem{nos} S. Pepin, Fl. Stancu, M. Genovese, J.-M.Richard and S. Zouzou,
work in progress.

\bibitem{12}L.Ya. Glozman and D.O. Riska, \Journal{{\em Phys. Rep.}}
{268}{263} {1996}. 
%
\bibitem{13}L.Ya. Glozman, Z. Papp and W. Plessas,
\Journal{\PLB} {381}{311} {1996}.                                          
%
\bibitem{Chromo} L. A. Copley, N. Isgur and G. Karl, \Journal{\PRD }
{20}{768} {1979}; S. Capstick and N. Isgur,\Journal{ \PRD} {34}
{2809} {1986};
C. S. Kalman and D. Pfeffer,\Journal{ \PRD} {27} {1648} {1983}.
%
\bibitem{14}N. T{\"o}rnqvist, \Journal{\PRL}{67} {556} {1991};
\Journal{\ZPC} {61} {525} {1994}.
%
\bibitem{15}A.V. Manohar and M.B. Wise, \Journal{\NPB} {399} {17} 
{1993}.
%
\bibitem{16}T.E.O. Ericson and G. Karl, \Journal{\PLB} {309}{ 426}
{1993}.
%
\bibitem{17}M. Weyrauch and H.J. Weber, \Journal{\PLB} {171} {13}
{1986};              
H.J. Weber and H.T. Williams, \Journal{\PLB} {205}{118}
{1988}.  

%
\bibitem{18}L.Ya. Glozman and D.O. Riska, \Journal{\NPA} {603}{ 326 }{1996}. 
%
\bibitem{buch} A. Buchman {\sl et al.}, \Journal{\NPA} {569}{ 661} 
{1994}.

\bibitem{Don} D. B. Lichtenberg, R. Roncaglia and E. Predazzi, 
IUHET-344, published in these proceedings.
\bibitem{19}{Fl. Stancu}, {Group Theory in Subnuclear Physics, Oxford University
Press, 1996,  chapter 4;}

\bibitem{PS} {S. Pepin and Fl. Stancu}{ Preprint ULG-PNT-96-1-J.}

\bibitem{Don2} D. B. Lichtenberg, R. Roncaglia and E. Predazzi, DFTT 
65/96, published in these proceedings.

\end{thebibliography}
\end{document}